%% file: Euphrosyne_final.tex
\documentclass[structabstract,longauth]{aa}
\usepackage{threeparttable,pifont}
\usepackage{natbib}
\usepackage{color}
\usepackage{multicol}
\usepackage{amsmath,amssymb,url}
\usepackage{graphicx}
\usepackage{xcolor}
\usepackage{txfonts,longtable,dcolumn,verbatim}
\bibpunct{(}{)}{;}{a}{}{,}
\usepackage{array} 
\usepackage[normalem]{ulem}

\usepackage[squaren,textstyle]{SIunits}

\usepackage{hyperref}
\hypersetup{
  bookmarks=true,         
    unicode=true,           
   pdftoolbar=true,        
    pdfmenubar=true,        
    pdffitwindow=true,      
    pdftitle={(31) Euphrosyne},
    pdfauthor={Yang et al.},
    pdfsubject={Planetary Science},
    pdfkeywords={},         
    pdfnewwindow=true,      
    colorlinks=true,        
    linkcolor=gray,         
    citecolor=blue,         
    filecolor=gray,         
    urlcolor=gray           
}

\usepackage{xspace}

\newcommand{\adam}{\texttt{ADAM}\xspace}
\newcommand{\genoid}{\texttt{Genoid}\xspace}
\newcommand{\mistral}{\texttt{Mistral}\xspace}

\newcommand{\Dens}{1\,665\,$\pm$\,242} 
\newcommand{\Diam}{268\,$\pm$\,6}
\newcommand{\Mass}{(1.7\,$\pm$\,0.3)\,$\times$\,$10^{19}$} 
\newcommand{\Masss}{1.7\,$\pm$\,0.3} 
\newcommand{\sid}{g$\cdot$cm$^{-3}$}
\newcommand{\sidd}{kg$\cdot$m$^{-3}$}

\title{Binary asteroid (31) Euphrosyne: Ice-rich and nearly spherical
    \thanks{Based on observations made with 
      ESO Telescopes at the La Silla Paranal Observatory under program 199.C-0074 (PI Vernazza)},\thanks{The  reduced  images  are  available  at  the  CDS  via  anonymous   ftp  to \url{http://cdsarc.u-strasbg.fr/} or via \url{http://cdsarc.u-strasbg.fr/viz-bin/qcat?J/A+A/xxx/Axxx}}}

\titlerunning{Physical model of (31)~Euphrosyne}

\author{B.~Yang\inst{\ref{eso}}            \and 
    J.~Hanu{\v s}\inst{\ref{prague}}       \and 
    B.~Carry\inst{\ref{oca}}               \and 
    P.~Vernazza\inst{\ref{lam}}            \and 
    M.~Bro\v{z}\inst{\ref{prague}}         \and 
    F.~Vachier\inst{\ref{imcce}}           \and 
    N.~Rambaux\inst{\ref{imcce}}           \and 
    M.~Marsset\inst{\ref{mit}}             \and 
    O.~Chrenko\inst{\ref{prague}}          \and 
    P.~\v{S}eve\v{c}ek\inst{\ref{prague}}  \and 
    M.~Viikinkoski\inst{\ref{tampere}}     \and 
    E.~Jehin\inst{\ref{liege}}             \and 
    M.~Ferrais\inst{\ref{lam}}           \and 
    E.~Podlewska-Gaca\inst{\ref{poznan}} \and 
    A.~Drouard\inst{\ref{lam}}             \and 
    F.~Marchis\inst{\ref{seti}}            \and 
    M.~Birlan\inst{\ref{imcce}, \ref{aira}} \and 
    Z. ~Benkhaldoun\inst{\ref{cau}} \and 
    J.~Berthier\inst{\ref{imcce}}          \and 
    P.~Bartczak\inst{\ref{poznan}}         \and 
    C.~Dumas\inst{\ref{tmt}}               \and 
    G.~Dudzi\'{n}ski\inst{\ref{poznan}}    \and
    J.~{\v D}urech\inst{\ref{prague}}      \and 
    J.~Castillo-Rogez\inst{\ref{jpl}}      \and 
    F.~Cipriani\inst{\ref{estec}}          \and 
    F.~Colas\inst{\ref{imcce}}             \and 
    R.~Fetick\inst{\ref{lam}}              \and 
    T.~Fusco\inst{\ref{lam},\ref{dota}}               \and 
    J.~Grice\inst{\ref{oca},\ref{ou}}      \and 
    L.~Jorda\inst{\ref{lam}}               \and 
    M.~Kaasalainen\inst{\ref{tampere}}     \and 
    A.~Kryszczynska\inst{\ref{poznan}}     \and 
    P.~Lamy\inst{\ref{lamos}}                \and 
    A.~Marciniak\inst{\ref{poznan}}        
    T.~Michalowski\inst{\ref{poznan}}      \and 
    P.~Michel\inst{\ref{oca}}              \and 
    M.~Pajuelo\inst{\ref{imcce},\ref{puc}} \and 
    T.~Santana-Ros\inst{{\ref{uda},\ref{iccub}}}      \and 
    P.~Tanga\inst{\ref{oca}}               \and 
    A.~Vigan\inst{\ref{lam}}               \and 
    O.~Witasse\inst{\ref{estec}}           
} 

   \institute{
     European Southern Observatory (ESO), Alonso de Cordova 3107, 1900 Casilla Vitacura, Santiago, Chile
     \label{eso}
    \and 
     Institute of Astronomy, Faculty of Mathematics and Physics, Charles University, V~Hole{\v s}ovi{\v c}k{\'a}ch 2, 18000 Prague, Czech Republic
     \label{prague}
     \and 
     Universit\'e C{\^o}te d'Azur, Observatoire de la C{\^o}te d'Azur, CNRS, Laboratoire Lagrange, France
     \label{oca}
      \and 
     Aix Marseille Univ, CNRS, LAM, Laboratoire d'Astrophysique de Marseille, Marseille, France
     \label{lam}
     \and 
     IMCCE, Observatoire de Paris, PSL Research University, CNRS, Sorbonne Universit{\'e}s, UPMC Univ Paris 06, Univ. Lille, France
     \label{imcce}
               \and 
      Department of Earth, Atmospheric and Planetary Sciences, MIT, 77 Massachusetts Avenue, Cambridge, MA 02139, USA
     \label{mit}
     \and
    Mathematics and Statistics, Tampere University, 33014 Tampere, Finland
     \label{tampere}
      \and 
     Space sciences, Technologies and Astrophysics Research Institute, Universit{\'e} de Li{\`e}ge, All{\'e}e du 6 Ao{\^u}t 17, 4000 Li{\`e}ge, Belgium
     \label{liege}
     \and 
     Astronomical Observatory Institute, Faculty of Physics, Adam Mickiewicz University, ul. S{\l}oneczna 36, 60-286 Pozna{\'n}, Poland
     \label{poznan}
       \and 
     SETI Institute, Carl Sagan Center, 189 Bernado Avenue, Mountain View CA 94043, USA 
     \label{seti}
     \and 
     Oukaimeden Observatory, High Energy Physics and Astrophysics Laboratory, Cadi Ayyad University, Marrakech, Morocco
     \label{cau}
      \and 
     Thirty-Meter-Telescope, 100 West Walnut St, Suite 300, Pasadena, CA 91124, USA
     \label{tmt}
     \and 
     Jet Propulsion Laboratory, California Institute of Technology, 4800 Oak Grove Drive, Pasadena, CA 91109, USA
     \label{jpl}
     \and 
     European Space Agency, ESTEC - Scientific Support Office, Keplerlaan 1, Noordwijk 2200 AG, The Netherlands
     \label{estec}
     \and 
     DOTA, ONERA, Université Paris Saclay, F-91123 Palaiseau, France
      \label{dota}
         \and 
     Open University, School of Physical Sciences, The Open University, MK7 6AA, UK
     \label{ou}
      \and 
     Laboratoire Atmosph\`eres, Milieux et Observations Spatiales, CNRS \& 
    Universit\'e de Versailles Saint-Quentin-en-Yvelines, Guyancourt, France
    \label{lamos}
     \and 
     Secci{\'o}n F{\'i}sica, Departamento de Ciencias, Pontificia Universidad Cat{\'o}lica del Per{\'u}, Apartado 1761, Lima, Per{\'u}
     \label{puc}
     \and 
     Departamento de Fisica, Ingenier\'ia de Sistemas y Teor\'ia de la Señal, Universidad de Alicante, Alicante, Spain
    \label{uda}    
      \and 
Institut de Ci\'encies del Cosmos (ICCUB), Universitat de Barcelona (IEEC-UB), Martí Franqu\'es 1, E08028 Barcelona, Spain
   \label{iccub}
     \and 
     Konkoly Observatory, Research Centre for Astronomy and Earth Sciences, Hungarian Academy of Sciences, Konkoly Thege 15-17, H-1121 Budapest, Hungary
     \label{konkoly}
     \and 
     Center for Solar System Studies, 446 Sycamore Ave., Eaton, CO 80615, USA
     \label{Birlan}
    \and 
     Astronomical Institute of Romanian Academy, 5, Cutitul de Argint Street, 040557 Bucharest, Romania
\label{aira}
}

   \date{Received x-x-2019 / Accepted x-x-2019}
 
  \abstract
   {Asteroid (31)~Euphrosyne is one of the biggest objects in the asteroid main belt and it is also the largest member of its namesake family. The Euphrosyne family occupies a highly inclined region in the outer main belt and contains a remarkably large number of members, which is interpreted as an outcome of a disruptive cratering event.}
  {The goals of this adaptive-optics imaging study were threefold: to characterize the shape of Euphrosyne, to constrain its density, and to search for the large craters that may be associated with the family formation event. } 
   {We obtained disk-resolved images of Euphrosyne using SPHERE/ZIMPOL at ESO's 8.2-m VLT as part of our large program (ID: 199.C-0074, PI: Vernazza). We reconstructed its 3D-shape using the \adam shape modeling algorithm based on the SPHERE images and the available lightcurves of this asteroid. We analyzed the dynamics of the satellite with the \genoid meta-heuristic algorithm. Finally, we studied the shape of Euphrosyne using hydrostatic equilibrium models.}
   {Our SPHERE observations show that Euphrosyne has a nearly spherical shape with the sphericity index of 0.9888 and its surface lacks large impact craters. Euphrosyne's diameter is 268$\pm$6 km, making it one of the top 10 largest main belt asteroids. We detected a  satellite of Euphrosyne -- S/2019 (31) 1-- that is about 4 km across, on an circular orbit. The mass determined from the orbit of the satellite together with the volume computed from the shape model imply a density of 1665$\pm$242 kg m$^{-3}$, suggesting that Euphrosyne probably contain a large fraction of water ice in its interior. We find that the spherical shape of Euphrosyne is a result of the reaccumulation process following the impact, as in the case of (10)~Hygiea. However, our shape analysis reveals that, contrary to Hygiea, the axis ratios of Euphrosyne significantly differ from the ones suggested by fluid hydrostatic equilibrium following reaccumulation.}
  

\begin{document}
  \maketitle

\section{Introduction}\label{sec:introduction}


The main asteroid belt is a dynamically living relic, with the shapes, sizes, and surfaces of most asteroids being altered by ongoing collisional fragmentation and cratering events \citep{Bottke2015}. Space probes and ground-based observations have revealed a fascinating variety among asteroid shapes, where large asteroids are nearly spherical \citep{Park2019,Vernazza2020} and small asteroids are irregularly shaped \citep{Durech2010,Shepard2017,Fujiwara2006,Thomas2012}. Most asteroids with diameters greater than $\sim$100 km have likely kept  their internal structure intact since their time of formation because the dynamical lifetime of those asteroids is estimated to be comparable to the age of the Solar system \citep{Bottke2005}. There are a few exceptions comprising essentially the largest remnants of giant families (e.g., (10)~Hygiea, \citealt{Vernazza2020}), whose shapes have been largely altered by the impact. In contrast, the shapes of smaller asteroids have been determined mainly through collisions, where their final shapes depend on collision conditions such as impact energies and spin rates \citep{Leinhardt2000,Sugiura2018}. 

The arrival of second generation extreme adaptive-optics (AO) instruments, such as the Spectro-Polarimetric High-contrast Exoplanet Research instrument (SPHERE) at VLT \citep{Beuzit2008} and the Gemini Planet Imager (GPI) at GEMINI-South \citep{Macintosh2014}, offers a great opportunity to study detailed shape, precise size and topographic feature of large main belt asteroids with diameter D$\ge$ 100 km via direct imaging.
AO-aided observations with high spatial resolution also enable detection of asteroidal satellites that are much smaller and closer to their primaries, which, thus far, could have remained undetected in prior searches \citep{Margot2015}. Consequently, physical properties that are not well constrained, such as the bulk density, the internal porosity and the surface tensile strength, can be investigated using AO corrected measurements. These are the key parameters that determine crater formation, family formation and/or satellite creation \citep{Michel2001}.

Asteroid (31)~Euphrosyne (hereafter, Euphrosyne) is the largest member of its namesake family. Previous studies have noted that the Euphrosyne family exhibits a very steep size frequency distribution (SFD), significantly depleted in large and medium sized asteroids \citep{Carruba2014}. Such a steep SFD is interpreted as a glancing impact between two large bodies resulting in a disruptive cratering event \citep{Masiero2015}. Euphrosyne is a Cb-type asteroid \citep{Bus2002} with an optical albedo of $p_V$ = 0.045$\pm$0.008 \citep{Masiero2013}. Euphrosyne's diameter has been reported as $D$ = 276$\pm$3 km \citep{Usui2011} or $D$ = 282$\pm$10 km \citep{Masiero2013} while its mass has been estimated by various studies leading to an average value of M$_{31}$=1.27$\pm$0.65 $\times 10^{19}$ kg, with about 50\% uncertainty \citep{Carry2012}. These size and mass estimates imply a density estimate of $\rho_{31}$=1180\,$\pm\,$610~\sidd{}. As detailed hereafter and was first reported in (CBET 4627, 2019), we discovered a satellite in this study, implying that it is one of the few large asteroids for which the density can be constrained with high precision \citep{Scheeres2015}.

In this paper, we present the high-angular resolution observations of Euphrosyne with VLT/SPHERE/ZIMPOL, which were obtained as part of our ESO large program (Sect.~\ref{sec:ao}). We use these observations together with a compilation of available and newly obtained optical lightcurves (Sect.~\ref{sec:lcs}) to constrain the 3D shape of Euphrosyne as well as its spin state and surface topography (Sect.~\ref{sec:ADAM}). We then describe the discovery of its small moonlet S/2019 (31) 1 (Sect.~\ref{sec:moon}) and constrain its mass by fitting the orbit of the satellite. Both the 3D shape (hence volume) and the mass estimate allow us to constrain the density of Euphrosyne with high precision (Sect.~\ref{sec:density}). We also use the AO images and the 3D shape model to search for large craters, which may be associated with the family-forming event. 


\section{Observations \& Data Reduction}\label{sec:data}

\subsection{Disk-resolved data with SPHERE}\label{sec:ao}

Euphrosyne was observed, between March and April 2019, using the Zurich Imaging Polarimeter (ZIMPOL) of SPHERE \citep{Thalmann2008} in the direct imaging mode with the narrow band filter (N$\_$R filter; filter central wavelength = 645.9 nm, width = 56.7 nm). The angular diameter of Euphrosyne was in the range of 0.16--0.17$\arcsec$ and the asteroid was close to an equator-on geometry at the time of the observations. Therefore, the SPHERE images of Euphrosyne obtained from seven epochs allow us to reconstruct a reliable 3D shape model with well defined dimensions. The reduced images were further deconvolved with the \mistral algorithm \citep{Fusco2003},
using a parametric point-spread function \citep{Fetick2019}.
Table~\ref{tab:ao} contains full information about the images. We display all obtained images in Fig.~\ref{fig:Deconv}.

\subsection{Disk-integrated optical photometry}\label{sec:lcs}

\begin{figure}
\begin{center}
\resizebox{1.0\hsize}{!}{\includegraphics{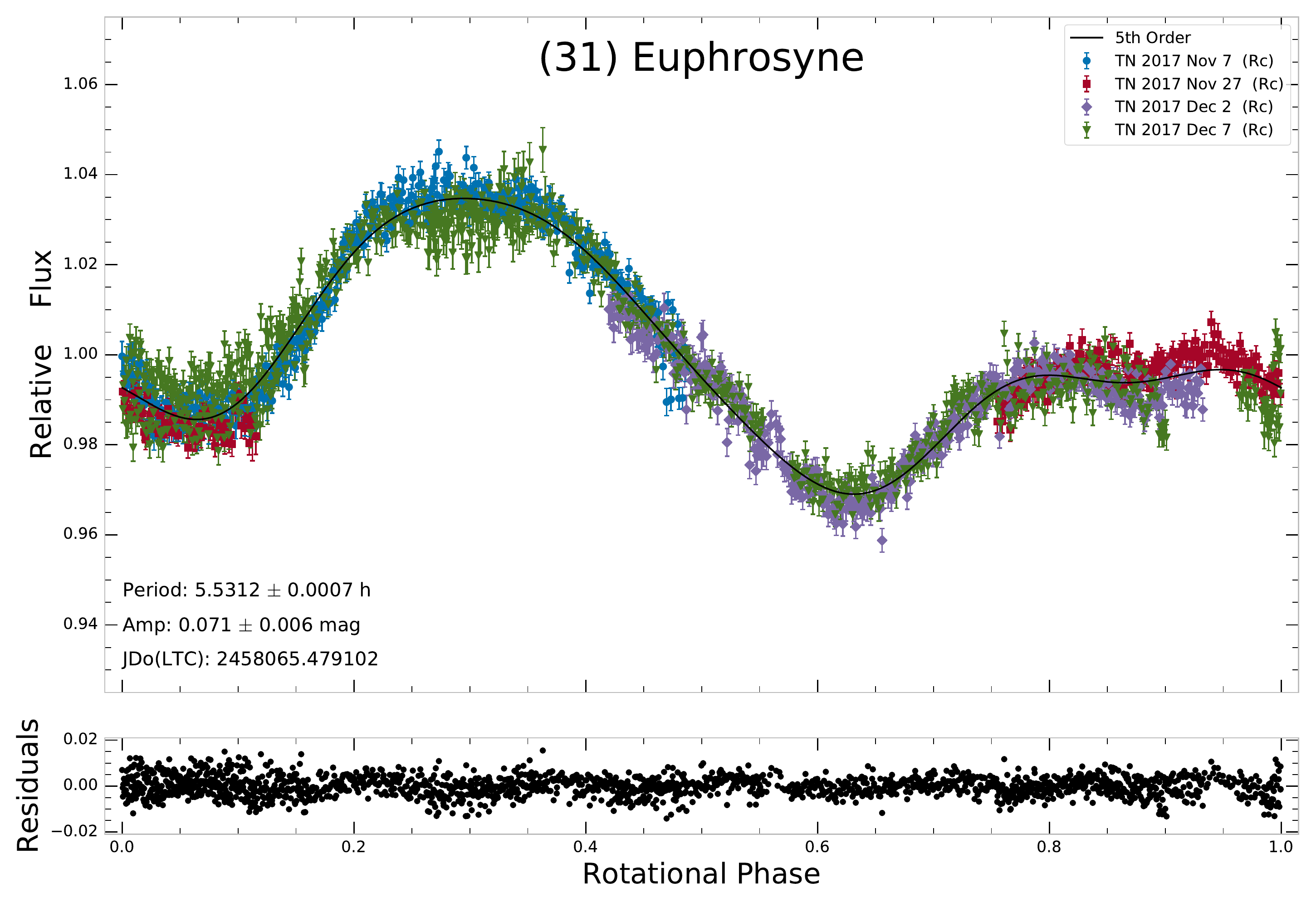}}\\
\end{center}
\caption{\label{fig:trappist} Composite lightcurve of (31)~Euphrosyne obtained with the TRAPPIST-North telescope in 2017 and a Fourier series
of fifth order is fitted to the data, shown as the solid line. The residuals of the fit are shown as black dots below. The details about the telescope and data format are described in \citep{Jehin2011}. }
\end{figure}

A set of 29 individual lightcurves of Euphrosyne was previously used by \citet{Hanus2016a} in order to derive a convex shape model of this large body. These lightcurves were obtained by \citet{Schober1980, Barucci1985, McCheyne1985, Kryszczynska1996, Pilcher2009b, Pilcher2012c}. We complemented these data with five additional lightcurves from the recent apparition in 2017: four lightcurves were obtained by the TRAPPIST North telescope (Fig.~\ref{fig:trappist}) and the fifth one was obtained via the Gaia-GOSA\footnote{\url{www.gaiagosa.eu}. It is a web-service platform that serves as a link between scientists seeking photometric data and amateur observers capable of obtaining such data with their small-aperture telescopes.}. Our final photometric dataset utilized for the shape modeling of Euphrosyne consists of 34 individual lightcurves. Detailed information about these lightcurves is provided in Table~\ref{tab:lcs}

\section{Determination of the 3D shape}\label{sec:ADAM}

\begin{figure*}
\begin{center}
\resizebox{0.99\hsize}{!}{\includegraphics{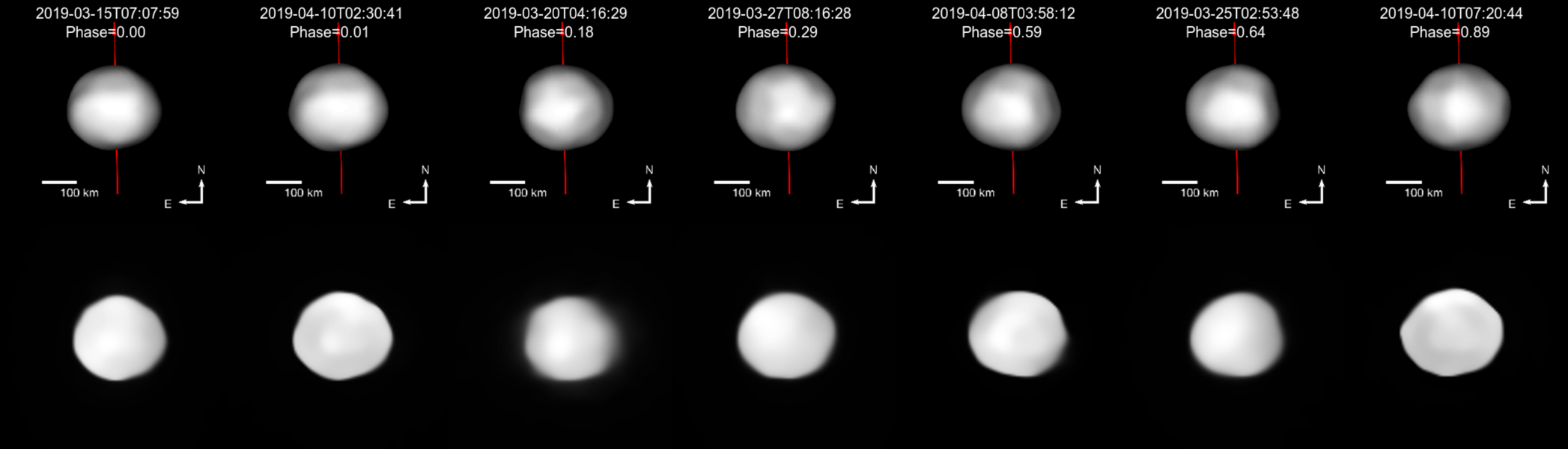}}\\
\end{center}
\caption{\label{fig:comparison}Comparison between the VLT/SPHERE/ZIMPOL deconvolved images of Euphrosyne (bottom) and the corresponding projections of our \adam{} shape model (top). The red line indicates the position of the rotation axis. We use a non-realistic illumination to highlight the local topography of the model.}
\end{figure*}

\begin{table}
 \caption{\label{tab:param}
 Physical properties of (31)~Euphrosyne based on \adam{} shape modeling and the \genoid{} orbit solution of the satellite: Sidereal rotation period $P$, spin-axis ecliptic J2000 coordinates $\lambda$ and $\beta$, volume-equivalent diameter $D$, dimensions along the major axis $a$, $b$, $c$, their ratios $a/b$ and $b/c$, mass $M$, and bulk density $\rho$. Uncertainties correspond to 1\,$\sigma$ values.}
 \centering
 \begin{threeparttable}
 \begin{tabular}{lll}
  \hline
  Parameter  & Unit & Value \\
  \hline
   $P$       & h    & 5.529595$\pm 10^{-6}$\\
   $\lambda$ & deg. & 94$\pm$5  \\
   $\beta$   & deg. & 67$\pm$3   \\
   $D$       & km   & \Diam  \\
   $a$       & km   & 294$\pm$6 \\
   $b$       & km   & 280$\pm$10  \\
   $c$       & km   & 248$\pm$6  \\
   $a/b$     &      & 1.05$\pm$0.03  \\
   $b/c$     &      & 1.13$\pm$0.04 \\
 $M$ & $10^{19}$~kg & \Masss \\
   $\rho$    & \sidd & \Dens \\
  \hline
 \end{tabular}
 \end{threeparttable}
\end{table}

The spin state solution from \citet{Hanus2016a} served as an initial input for the shape modeling with the All-data Asteroid Modeling algorithm \citep[{\adam},][]{Viikinkoski2015} that fits simultaneously the optical data and the disk-resolved images. We followed the same shape modeling approach applied in our previous studies based on disk-resolved data from the SPHERE large program \citep[for instance,][]{Vernazza2018, Viikinkoski2018, Hanus2019a}. First, we constructed a low resolution shape model based on all available data, then we used this shape model as a starting point for further modeling with decreased weight of the lightcurves and increased shape model resolution. We performed this approach iteratively until we were satisfied with the fit to the lightcurve and disk-resolved data. We also tested two different shape parametrizations -- octantoids and subdivision \citep{Viikinkoski2015}. We show the comparison between the VLT/SPHERE/ZIMPOL deconvolved images of Euphrosyne and the corresponding projections of the shape model in Fig.~\ref{fig:comparison}.

\setkeys{Gin}{draft=false}
\begin{figure}
\resizebox{1.0\hsize}{!}{\includegraphics{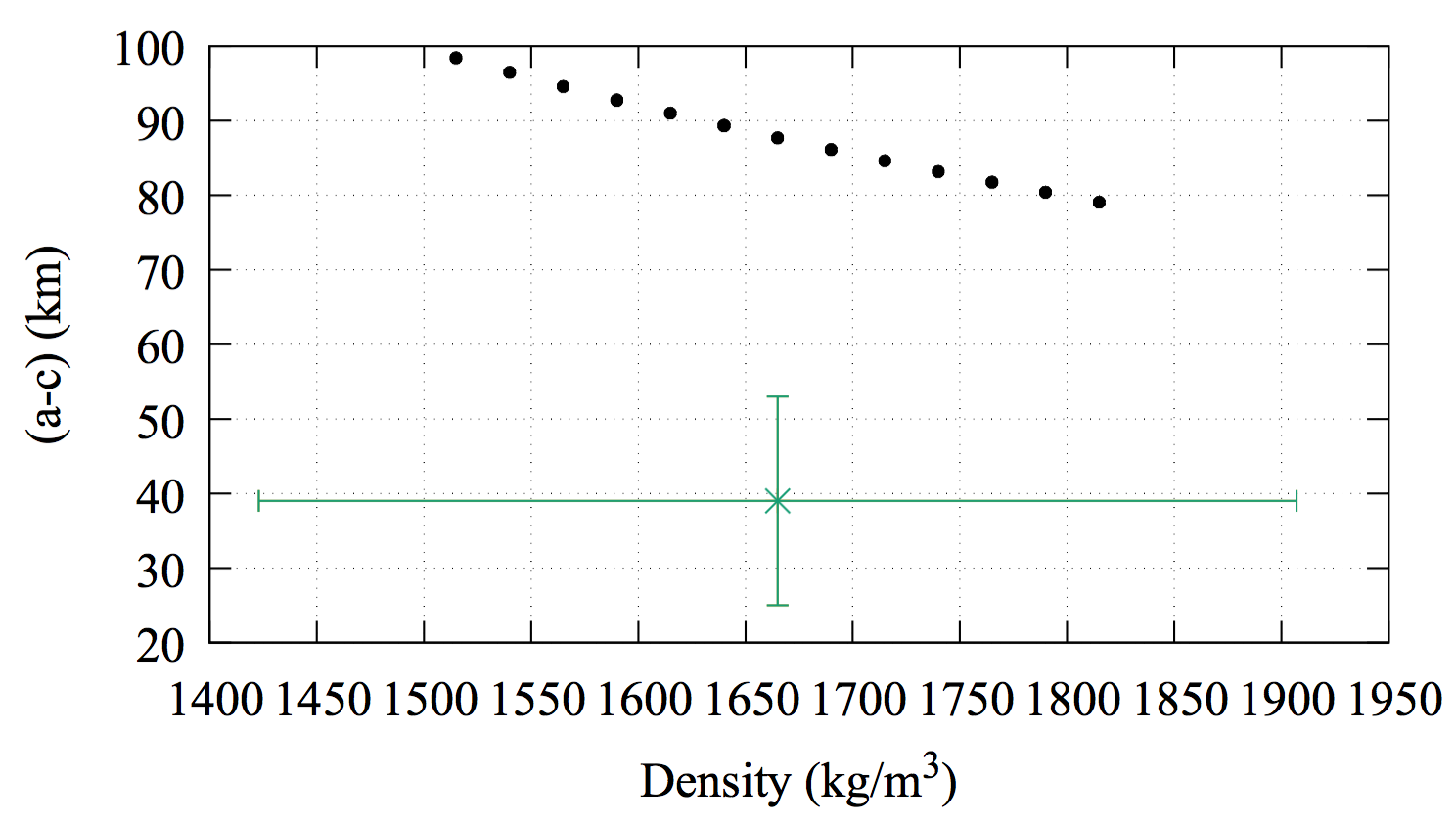}}
\caption{The calculated (a - c) for (31)~Euphrosyne as function of mean density for homogeneous case, given Euphrosyne's rotation period of 5.53h, are shown as black dots. The green star represents the value derived in Sect. \ref{sec:density} with its 1-$\sigma$ uncertainty (uncertainties of a and c are added quadratically).}
\label{fig:shape}
\end{figure}
\setkeys{Gin}{draft=true}

Owing to the nearly equator-on geometry of the asteroid, our images taken at seven different rotation phases have a nearly complete coverage of the entire surface of Euphrosyne. The SPHERE data enable an accurate determination of Euphrosyne's dimensions, including the ones along the rotation axis. The physical properties of Euphrosyne derived are listed in Table~\ref{tab:param}. The uncertainties reflect the dispersion of values obtained with various shape models based on different data weighting, shape resolution and parametrization. These values correspond to about 1 pixel, which is equivalent to $\sim$ 5.93~km. Our volume equivalent diameter $D$=\Diam~km is consistent within 1$\sigma$ with the radiometric estimates of \citet[$D$=276$\pm$3~km,][]{Usui2011} and \citet[$D$=282$\pm$10~km,][]{Masiero2013}. The shape of Euphrosyne is fairly spherical with almost equal equatorial dimensions ($a/b$=1.05$\pm$0.03) and only a small flattening ($b/c$=1.13$\pm$0.04) along the spin axis. Euphrosyne's sphericity index (see \citealt{Vernazza2020} for more details) is equal to 0.9888, which is somewhat higher than that of (4)~Vesta, (2)~Pallas and (704)~Interamnia \citep{Vernazza2020, Hanus2020} making it so far the 3rd most spherical main belt asteroid after Ceres and Hygiea. 

Given the rather spherical shape of Euphrosyne and the fact that its a and b axes have similar lengths (within errors), we investigated whether the shape of Euphrosyne is close to hydrostatic equilibrium, using the same approach as described in \citep{Hanus2020}. It appears that Euphrosyne's shape is significantly different from the Maclaurin spheroid, which would be much flatter along the c axis (see Fig. \ref{fig:shape}).  Our SPHERE observations show that Euphrosyne is not actually in hydrostatic equilibrium for its current rotation, which will be further discussed in Sect.~\ref{sec:discuss}.

\begin{table}[!h]
\begin{center}\caption[Orbital elements of the satellite of Euphrosyne]{Orbital elements of the satellite of Euphrosyne, expressed in EQJ2000, obtained with \genoid: orbital period $P$, semi-major axis $a$, eccentricity $e$, inclination $i$, longitude of the ascending node $\Omega$, argument of pericenter $\omega$, time of pericenter $t_p$. The number of observations and RMS between predicted and observed positions are also provided. Finally, we report the mass of Euphrosyne $M_{\textrm{Euphrosyne}}$, the ecliptic J2000 coordinates of the orbital pole ($\lambda_p,\,\beta_p$), the equatorial J2000 coordinates of the orbital pole ($\alpha_p,\,\delta_p$), and the orbital inclination ($\Lambda$) with respect to the equator of Euphrosyne. Uncertainties are given at 3-$\sigma$.}
  \label{tab:dyn} 
   \begin{tabular}{l ll}
    \hline\hline
    & \multicolumn{2}{c}{S2019-31-1}\\ 
    \hline
  \noalign{\smallskip}
  \multicolumn{2}{c}{Observing data set} \\
  \noalign{\smallskip}
    Number of observations  & \multicolumn{2}{c}{5} \\ 
    Time span (days)        & \multicolumn{2}{c}{26} \\ 
    RMS (mas)               & \multicolumn{2}{c}{1.52} \\ 
    \hline
  \noalign{\smallskip}
  \multicolumn{2}{c}{Orbital elements EQJ2000} \\
  \noalign{\smallskip}
    $P$ (day)         & 1.209 & $\pm$ 0.003 \\ 
    $a$ (km)          & 672 & $\pm$ 35 \\ 
    $e$               & 0.043 & $_{-0.043}^{+0.123}$ \\ 
    $i$ (\degr)       & 1.4 & $\pm$ 1.4 \\ 
    $\Omega$ (\degr)  & 80.1 & $\pm$ 27.9 \\ 
    $\omega$ (\degr)  & 135.2 & $\pm$ 40.5 \\ 
    $t_{p}$ (JD)      & 2458565.33 & $\pm$ 0.13 \\ 
    \hline
  \noalign{\smallskip}
  \multicolumn{2}{c}{Derived parameters} \\
  \noalign{\smallskip}
    $M_{\textrm{Euphrosyne}}$ ($\times 10^{19}$ kg)      & 1.648 & $\pm$ 0.264 \\ 
    $\lambda_p,\,\beta_p$ (\degr)  & 86, +67 & $\pm$ 3, 2 \\ 
    $\alpha_p,\,\delta_p$ (\degr)  & 350, +89 & $\pm$ 21, 3 \\ 
    $\Lambda$ (\degr)              & 1 & $\pm$  2 \\ 
    \hline
  \end{tabular}
\end{center}
\end{table}

\section{Orbital properties of the satellite}\label{sec:moon}

\setkeys{Gin}{draft=false}
\begin{figure*}
\includegraphics[width=\textwidth]{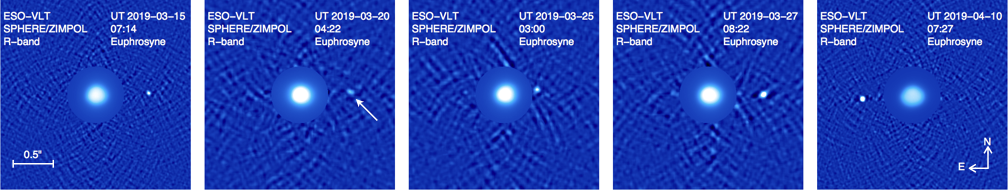}
\caption{Processed ZIMPOL images, revealing the presence of the satellite, S/2019 (31) 1, around (31)~Euphrosyne in five epochs.  The pixel intensities within 0.22$''$ of the primary have been reduced by a factor of $\sim$ 2000 to increase the visibility of the faint satellite. The images were smoothed by convolving a Gaussian function with FWHM of $\sim$ 8 pixels. The arrow points out the location of the satellite in the image taken on UT March 20, 2019, when the satellite appeared very dim compared to the other nights.}
\label{figsat}
\end{figure*} 
\setkeys{Gin}{draft=true}

Each image obtained with SPHERE/ZIMPOL was further processed to remove the bright halo surrounding Euphrosyne, following the procedure described in details in (\citealt{Pajuelo2018} and \citealt{Yang2016}). The residual structures after the halo removal were minimized using the processing techniques introduced in \citep{Wahhaj2013}, where the background structures were removed using a running median in a $\sim$ 50 pixel box in the azimuthal direction as well as in a $\sim$ 40 pixel box in the radial direction. Adopting the method introduced in \citet{Yang2016}, we inserted 100 point sources, which known intensity and FWHM, in each science image to estimate flux loss due to the halo removal processes. In five out of of seven epochs, a faint non-resolved source was clearly detected in the vicinity of Euphrosyne (Fig~\ref{figsat}). The variation in the brightness of the satellite is mainly due to the difference in the atmospheric conditions at the time of the observations, which directly affect the AO performance.

We measured the relative positions on the plane of the sky between Euphrosyne and its satellite \citep[fitting two 2D Gaussians, see][]{Carry2019} and report them in Table~\ref{tab:genoid}. We then used the \genoid~algorithm \citep{Vachier2012} to determine the orbital elements of the satellite. The best solution fits the observed positions with root mean square (RMS) residuals of 1.5\,mas only (Table~\ref{tab:dyn}). The orbit of the satellite is circular, prograde, and equatorial, similar to most known satellites around large main belt asteroids \citep[e.g.,][]{Marchis2008,Berthier2014,Margot2015,Carry2019}.

\setkeys{Gin}{draft=false}
\begin{figure}
\includegraphics[width=4.3cm]{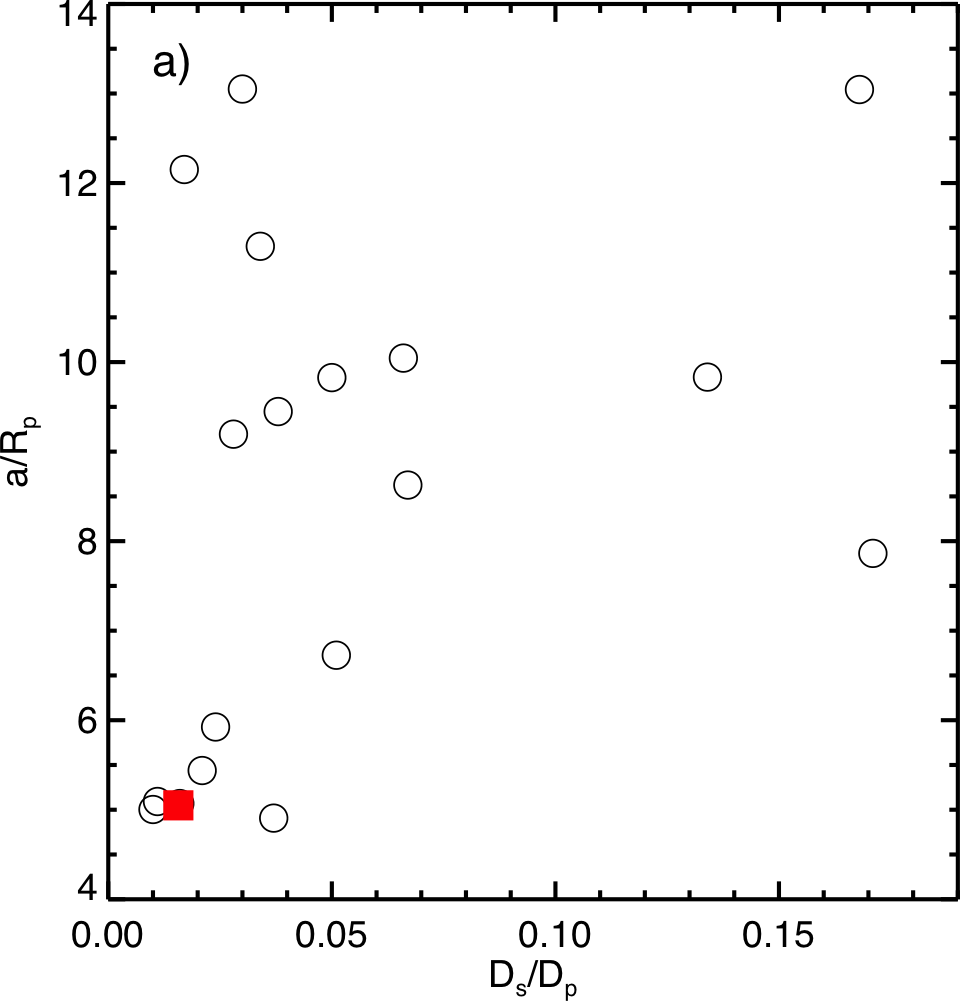} \hspace{0.1cm} \includegraphics[width=4.3cm]{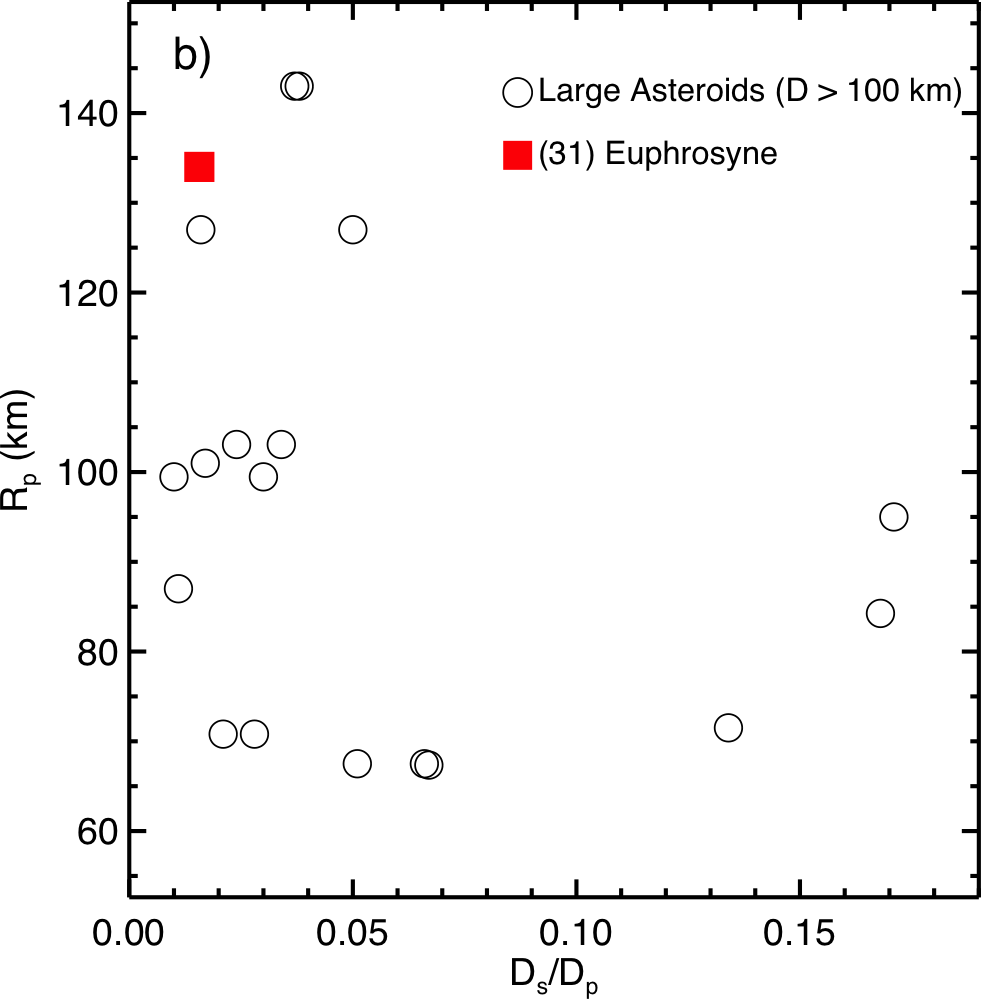}
\caption{\label{sate_comp} a). Relative component separation and b). primary radius versus secondary to primary diameter ratio for presently known main belt binary/triple asteroids \citep{Johnston2018}. Large asteroids with diameter greater than 100 km are shown as open circles. Asteroid (90) Antiope is excluded because of its unusually large secondary to primary diameter ratio. Our measurements of (31)~Euphrosyne are shown as the red squares. }   
\end{figure} 
\setkeys{Gin}{draft=true}

From the difference in the apparent magnitude of 9.0\,$\pm$\,0.3 between Euphrosyne and S/2019 (31) 1, and assuming a similar albedo for both, we estimate the diameter of the satellite to be 4.0\,$\pm$\,1.0\,km. The Euphrosyne binary system has the relative component separation $a/R_P$ = 5.0 $\pm$ 0.3 and the secondary-to-primary diameter ratio $D_s/D_p$=0.015 $\pm$ 0.005, where $a$ is the semi-major axis of the system, $D_s$ and $D_p$ are the diameter of the satellite and the primary respectively and $R_P$ is the radius of the primary. The comparison of the properties of the Euphrosyne binary system to other large asteroid systems are shown in Fig. \ref{sate_comp}. Compared to the other system, S/2019 (31) 1 has one of the smallest secondary-to-primary diameter ratios and is very close to the primary. Given the small size of the satellite, S/2019 (31) is expected to be tidally locked, i.\ e.\ its spin  period synchronizes to its orbital period on million year timescale \citep{Rojo2011}. 

\section{Bulk density and surface topography}\label{sec:density}
Owing to the presence of the satellite, we derived the mass of the system \Mass\,kg with a fractional precision of 15\%, which is considerably better than all the previous indirect measurements. Combining our mass measurement with the newly derived volume based on our 3D-shape, we obtain a bulk density of 1.665$\pm$0.242~\sid\ for Euphrosyne.

We note that the bulk density of Euphrosyne is the lowest among all the other large C-type asteroids measured to date, e.g. (1)~Ceres \citep[2.161$\pm$0.003 \sid, D$\sim$1000 km,][]{Park2019}, (10)~Hygiea \citep[1.94$\pm$0.25 \sid, D$\sim$440 km,][]{Vernazza2020} and (704)~Interamnia \citep[1.98$\pm$0.68 \sid, D$\sim$300km,][]{Hanus2020}. On the other hand, such density around 1.7 \sid or lower is more common among intermediate sized C-type asteroids, such as (45)~Eugenia \citep[1.4$\pm$0.4 \sid, D$\sim$200 km,][]{Marchis2012b}, (93)~Minerva \citep[1.75$\pm$0.68 \sid, D$\sim$160 km,][]{Marchis2013a}, (130)~Eletra \citep[1.60$\pm$0.13 \sid, D$\sim$200 km,][]{Hanus2016c} and (762) Pulcova \citep[0.8$\pm$0.1 \sid, D$\sim$150 km,][]{Marchis2012b}.


For the C-complex asteroids mentioned above, the density seems to show a trend with size, where the smaller asteroids have lower densities. Such trend could be explained by increasing porosity in smaller asteroids. Nonetheless, as already discussed in recent works \citep{Carry2012,Viikinkoski2015b, Marsset2017, Carry2019, Hanus2020}, the macroporosity of Euphrosyne is likely to be small ($\leq$20\%) due to its relatively high internal pressure owing to its large mass $>$ 10$^{19}$kg. Given the small macroporosity of Euphrosyne, its density, therefore, is diagnostic of its bulk composition. As for the other large C-type asteroids (Ceres, Hygiea, Interamnia), a large amount of water must be present in Euphrosyne. Assuming 20\% porosity, a typical density of anhydrous silicates of 3.4 g/cm$^3$ and a density of 1.0 g/cm$^3$ for water ice, the presence of water ice, up to 50\% by volume, is required in the interior of Euphrosyne to match its bulk density. 

In terms of topographic characteristics, the surface of Euphrosyne appears smooth and nearly featureless without any apparent large basins. This is in contrast to other objects studied by our large program that show various-sized craters on their surfaces, such as (2) Pallas \citep[B-type,][]{Marsset2020}, (4)~Vesta \citep[V-type,][]{Fetick2019} and (7)~Iris \citep[S-type,][]{Hanus2019a}. On the other hand, lacking surface features in AO images is not unprecedented among large asteroids, especially among C-type asteroids. Ground-based AO observations have identified at least three other cases that are lacking prominent topographic structures, namely (1)~Ceres
\citep{Carry2008}, (10)~Hygiea \citep{Vernazza2020}, and (704)~Interamnia \citep{Hanus2020}. Although NASA/Dawn observations revealed a highly cratered surface of Ceres \citep{Hiesinger2016}, this dwarf planet clearly lacks large craters which suggests rapid viscous relaxation or protracted resurfacing due to the presence of large amounts of water \citep{Marchi2016}.

Similarly, as discussed already for the cases of Hygiea \citep{Vernazza2020} and Interamnia \citep{Hanus2020}, the absence of apparent craters in our Euphrosyne images may be due to the flat-floored shape of D$\geq$40km craters (this diameter corresponds to the minimum size of features that can be recognized on the surface of Euphrosyne), which would be coherent with a high water content for this asteroid, in agreement with our bulk density estimate. 



\section{Discussion}\label{sec:discuss}

In an attempt to understand the unexpected nearly spherical shape of Euphrosyne, we adopted a similar approach as described in \citet{Vernazza2020} and used hydrodynamical simulations to study the family-formation event. The simulations were performed with a smoothed- particle hydrodynamics (SPH) code to constrain the impact parameters, such as the impact angle and the diameter of the impactor.  We assumed the target and the impactor are both monolithic bodies with an initial density of the material $\rho_0$ = 1.665 \sid, corresponding to the present-day density of Euphrosyne. Our SPH simulations find that the impact event of Euphrosyne is even more energetic in comparison to that of Hygiea. As such, the parent body of Euphrosyne is completely fragmented by the impact and the final reaccumulated shape of Euphrosyne is highly spherical, which is similar to the case of Hygiea where the nearly round shape is formed following post-impact reaccumulation \citep{Vernazza2020}. 
  
We further studied the orbital evolution of the Euphrosyne family and determined the age of the family, using the newly developed method \citep{Broz_Morbidelli_2019Icar..317..434B}. 
Our N-body simulations further constrained the age of the Euphrosyne family to $\tau \sim $280 Myr that is significantly younger than the previous estimates \citep[between 560 and 1160 Myr,][]{Carruba2014}. The details of our SPH simulation as well as a full characterization of the Euphrosyne family are presented in a forthcoming article (Yang et al., in press). The young dynamical age and post-impact re-accumulation, collectively, may have contributed to the apparent absence of craters on the surface of Euphrosyne. 

Our new finding about the young age of Euphrosyne makes this asteroid a unique object for us to study the impact aftermath on a very young body that is only ${\sim}\,0.3\,{\rm Gyr}$ old. Previously, the SPH simulations for the case of Hygiea showed that its shape relaxed to a sphere during the gravitational reaccumulation phase, accompanied by an acoustic fluidization. The relaxation process on Hygiea could have settled down on a timescale of a few hours \citep{Vernazza2020}. However, the shape relaxation, in theory, maybe a rather long-term process, which could possibly last as long as the age of the body ($\tau$= 3\,Gyr, as suggested by its family).  If the physical mechanisms work the same way on both bodies, then the relaxation timescale simply can not be short on one body ($D = 268\,{\rm km}$) and  be 10 times longer on the other, larger, one ($D = 434\,{\rm km}$). To reconcile with both observations, the shape relaxation, if it is a long-term process, should occur on timescales that are comparable to or less than $0.3\,{\rm Gyr}$.

In addition to the much younger dynamical age, the rotation period of Euphrosyne is also shorter (P=5.53 h) than those of Hygiea and Ceres. As noted in \citet{Descamps2011}, the spin rate of Euphrosyne is faster than the typical rotation rates of asteroids with similar sizes. This is interpreted as a result of a violent disruption process, where the parent body is re-accumulated into high angular momentum shape and spin configuration \citep{Walsh2015}. With that spin rate, we would expect Euphrosyne to have a shorter $c$ axis compared to the $a$ axis using MacLaurin's equation \citep{Chandrasekhar1969} as shown in Fig. \ref{fig:shape}. However, a MacLaurin ellipsoid represents the hydrostatic equilibrium figure of a homogeneous and intact body, which is not the case for Euphrosyne since it is a re-accumulated body. This may explain why the actual shape of Euphrosyne deviates from that of a MacLaurin ellipsoid. 


\section{Conclusions}\label{sec:results}
In this paper, we presented high angular imaging observations of asteroid (31)~Euphrosyne and its moon. Our main findings are summarized as follows:\\
\noindent 1). The disk-resolved images and the 3D-shape model of Euphrosyne show that it is the third most spherical body among the main belt asteroids with known shapes after Ceres and Hygiea. Its round shape is consistent with a re-accumulation event following the giant impact at the origin of the Euphrosyne family. 

\noindent 2). The orbit of Euphrosyne's satellite, S/2019 (31) 1, is circular, prograde, and equatorial, similar to most known satellites around large main belt asteroids. The estimated diameter of this newly detected satellite is 4$\pm$1 km, assuming a similar albedo for the satellite and the primary. 

\noindent 3). The bulk density of Euphrosyne is 1665$\pm$242~\sidd, which is the first high precision density measurement via ground-based  observations for a Cb type asteroid.  Such density implies that a large amount of water (at least 50\% in volume) must be present in Euphrosyne.

\noindent 4). The surface of Euphrosyne is nearly featureless with no large craters detected, which is consistent with its young age and ice-rich composition.

\begin{acknowledgements}
This work has been supported by the Czech Science Foundation through grant 18-09470S (J. Hanu\v s, J. \v Durech, O. Chrenko, P. \v Seve\v cek) and by the Charles University Research program No. UNCE/SCI/023. Computational resources were supplied by the Ministry of Education, Youth and Sports of the Czech Republic under the projects CESNET (LM2015042) and IT4Innovations National Supercomputing Centre (LM2015070). P.~Vernazza, A.~Drouard, M. Ferrais and B.~Carry were supported by CNRS/INSU/PNP.  M.M. was supported by the National Aeronautics and Space Administration under grant No. 80NSSC18K0849 issued through the Planetary Astronomy Program. The work of TSR was carried out through grant APOSTD/2019/046 by Generalitat Valenciana (Spain). This work was supported by the MINECO (Spanish Ministry of Economy) through grant RTI2018-095076-B-C21 (MINECO/FEDER, UE). The research leading to these results has received funding from the ARC grant for Concerted Research Actions, financed by the Wallonia-Brussels Federation. TRAPPIST is a project funded by the Belgian Fonds (National) de la Recherche Scientifique (F.R.S.-FNRS) under grant FRFC 2.5.594.09.F. TRAPPIST-North is a project funded by the Universit{\'e} de Li{\`e}ge, and performed in collaboration with Cadi Ayyad University of Marrakesh. E. Jehin is a FNRS Senior Research Associate.

\end{acknowledgements}

\newpage

\begin{appendix}

\section{Additional figures and tables}

\setkeys{Gin}{draft=false}
\begin{figure*}[!h]
\begin{center}
\resizebox{0.8\hsize}{!}{\includegraphics{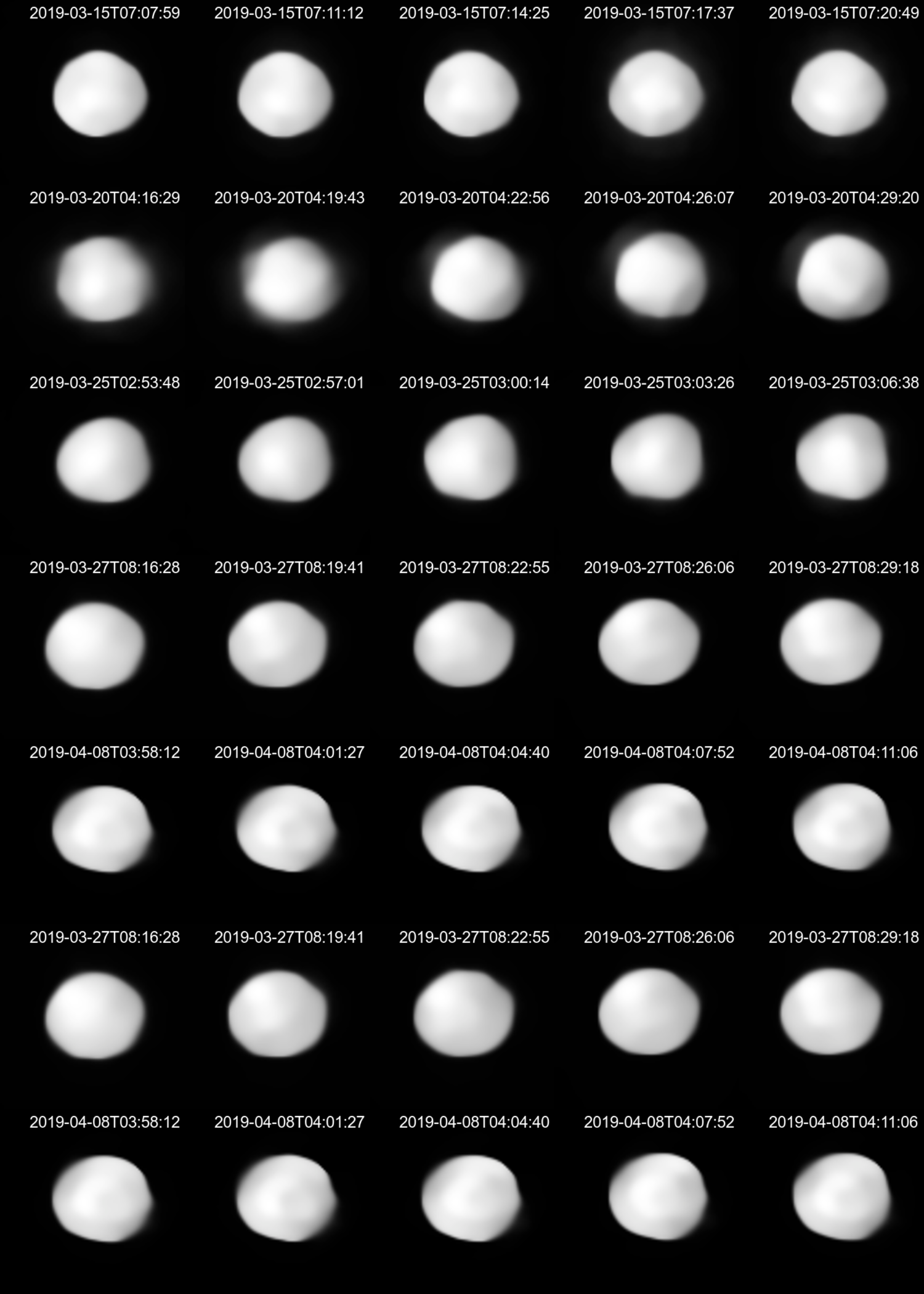}}\\
\end{center}
\caption{\label{fig:Deconv}Full set of VLT/SPHERE/ZIMPOL images of (31)~Euphrosyne. We show the images deconvolved by the \mistral~algorithm. Table~\ref{tab:ao} contains full information about the data.}
\end{figure*}
\setkeys{Gin}{draft=true}

\input{tab_ao.tex}

\begin{table*}
\caption{\label{tab:lcs}List of optical disk-integrated lightcurves used for \adam{} shape modeling. For each lightcurve, the table gives the epoch, the number of individual measurements $N_p$, asteroid's distances to the Earth $\Delta$ and the Sun $r$, phase angle $\varphi$, photometric filter and observation information.}
\begin{tabular}{rlr rrr l l}
\hline 
\hline
\multicolumn{1}{c} {N} & \multicolumn{1}{c} {Epoch} & \multicolumn{1}{c} {$N_p$} & \multicolumn{1}{c} {$\Delta$} & \multicolumn{1}{c} {$r$} & \multicolumn{1}{c} {$\varphi$} & \multicolumn{1}{c} {Filter} & Reference \\
 &  &  & (AU) & (AU) & (\degr) &  &  \\
\hline
     1  &  1977-09-24.3  &  41   &  2.18  &  3.13  &  7.4   &  V  &  \citet{Schober1980}       \\
     2  &  1978-11-14.0  &  95   &  1.84  &  2.43  &  21.7  &  V  &  \citet{Schober1980}      \\
     3  &  1978-11-16.0  &  74   &  1.83  &  2.43  &  21.4  &  V  &  \citet{Schober1980}      \\
     4  &  1978-11-19.0  &  78   &  1.80  &  2.43  &  21.0  &  V  &  \citet{Schober1980}      \\
     5  &  1979-01-01.4  &  16   &  1.61  &  2.44  &  15.3  &  V  &  \citet{Schober1980}      \\
     6  &  1983-10-29.0  &  81   &  1.72  &  2.71  &  2.0   &  V  &  \citet{Barucci1985}      \\
     7  &  1983-11-24.9  &  25   &  1.81  &  2.65  &  13.3  &  V  &  \citet{McCheyne1985}      \\
     8  &  1983-11-25.9  &  15   &  1.82  &  2.65  &  13.7  &  V  &  \citet{McCheyne1985}      \\
     9  &  1994-10-31.9  &  100  &  1.90  &  2.81  &  9.4   &  R  &  \citet{Kryszczynska1996}  \\
    10  &  2008-04-06.4  &  202  &  2.43  &  3.37  &  6.8   &  R  &  \citet{Pilcher2009b}       \\
    11  &  2008-04-10.3  &  246  &  2.42  &  3.38  &  5.5   &  R  &  \citet{Pilcher2009b}     \\
    12  &  2008-04-15.3  &  213  &  2.40  &  3.39  &  3.8   &  R  &  \citet{Pilcher2009b}     \\
    13  &  2008-04-25.3  &  224  &  2.40  &  3.41  &  1.0   &  R  &  \citet{Pilcher2009b}     \\
    14  &  2009-06-05.7  &  72   &  2.95  &  3.85  &  7.9   &  R  &  \citet{Pilcher2009b}     \\
    15  &  2009-06-07.6  &  85   &  2.95  &  3.85  &  7.8   &  R  &  \citet{Pilcher2009b}     \\
    16  &  2009-06-09.6  &  51   &  2.94  &  3.85  &  7.6   &  R  &  \citet{Pilcher2009b}     \\
    17  &  2009-06-10.7  &  110  &  2.94  &  3.85  &  7.6   &  R  &  \citet{Pilcher2009b}     \\
    18  &  2009-06-11.6  &  141  &  2.94  &  3.85  &  7.5   &  R  &  \citet{Pilcher2009b}     \\
    19  &  2009-06-22.5  &  12   &  2.94  &  3.86  &  7.6   &  R  &  \citet{Pilcher2009b}     \\
    20  &  2011-09-28.0  &  200  &  1.87  &  2.69  &  14.7  &  C  &  \citet{Hanus2016a}     \\
    21  &  2011-10-03.0  &  179  &  1.82  &  2.69  &  13.2  &  C  &  \citet{Hanus2016a}          \\
    22  &  2011-10-11.3  &  285  &  1.75  &  2.67  &  10.3  &  C  &  \citet{Pilcher2012c}     \\
    23  &  2011-11-01.4  &  287  &  1.64  &  2.63  &  2.8   &  C  &  \citet{Pilcher2012c}     \\
    24  &  2011-11-16.3  &  282  &  1.65  &  2.61  &  6.9   &  C  &  \citet{Pilcher2012c}     \\
    25  &  2011-12-10.2  &  378  &  1.78  &  2.57  &  15.9  &  C  &  \citet{Pilcher2012c}     \\
    26  &  2013-01-28.4  &  278  &  2.12  &  2.77  &  17.5  &  R  &  \citet{Pilcher2012c}     \\
    27  &  2013-02-20.4  &  300  &  1.97  &  2.82  &  12.3  &  R  &  \citet{Pilcher2012c}     \\
    28  &  2013-02-25.4  &  396  &  1.95  &  2.83  &  11.2  &  R  &  \citet{Pilcher2012c}     \\
    29  &  2013-04-17.3  &  343  &  2.13  &  2.94  &  13.6  &  R  &  \citet{Pilcher2012c}     \\
    30  &  2017-11-8.1   &  359  &  1.79  &  2.47  &  19.9  &  R  &  E. Jehin, M. Ferrais, Trappist North  \\
    31  &  2017-11-28.1  &  261  &  1.66  &  2.46  &  16.3  &  R  &  E. Jehin, M. Ferrais, Trappist North  \\
    32  &  2017-12-3.1   &  365  &  1.63  &  2.46  &  15.5  &  R  &  E. Jehin, M. Ferrais, Trappist North  \\
    33  &  2017-12-8.3   &  864  &  1.62  &  2.46  &  14.8  &  R  &  E. Jehin, M. Ferrais, Trappist North  \\
    34  &  2018-2-27.9   &  244  &  2.03  &  2.49  &  22.6  &  C  &  Gaia-GOSA                 \\
   \hline
\end{tabular}
\tablefoot{
     Gaia-GOSA (Gaia-Ground-based Observational Service for Asteroids, \url{www.gaiagosa.eu}).
    }
\end{table*}

\section{Astrometry of the satellite}
\onecolumn
\begin{center}
  \begin{longtable}{cclclrrrrrrr}
  \caption[Astrometry of S/2019 (31) 1]{Astrometry of Euphrosyne's satellite S/2019 (31) 1.
    Date, mid-observing time (UTC), telescope, camera, filter, 
    astrometry ($X$ is aligned with Right Ascension, and $Y$ with Declination,
    $o$ and $c$ indices stand for observed and computed positions, and $\sigma$ is pixel scale),
    and photometry (magnitude difference $\Delta M$ with uncertainty $\delta M$).
    \label{tab:genoid}
  }\\
    \hline\hline
     Date & UTC & Tel. & Cam. & Filter &
     \multicolumn{1}{c}{$X_o$} &
     \multicolumn{1}{c}{$Y_o$} &
     \multicolumn{1}{c}{$X_{o-c}$} &
     \multicolumn{1}{c}{$Y_{o-c}$} &
     \multicolumn{1}{c}{$\sigma$} &
     \multicolumn{1}{c}{$\Delta M$} &
     \multicolumn{1}{c}{$\delta M$} \\
    &&&&& 
     \multicolumn{1}{c}{(mas)} & \multicolumn{1}{c}{(mas)} & 
     \multicolumn{1}{c}{(mas)} & \multicolumn{1}{c}{(mas)} & 
     \multicolumn{1}{c}{(mas)} & 
     \multicolumn{1}{c}{(mag)} & \multicolumn{1}{c}{(mag)}  \\ 
    \hline
    \endhead
    \hline
    \endlastfoot
2019-03-15 & 07:07:59.26 & VLT & SPHERE/ZIMPOL & R & -398.3 & 10.5 & -0.7 & 2.1 & 3.6 & 8.9 & 0.2  \\
2019-03-20 & 04:16:29.04 & VLT & SPHERE/ZIMPOL & R & -377.7 & 12.6 & 1.4 & -3.4 & 3.6 & 9.2 & 0.4  \\
2019-03-25 & 02:53:48.81 & VLT & SPHERE/ZIMPOL & R & -239.7 & 30.7 & -0.6 & 0.9 & 3.6 & 9.1 & 0.3  \\
2019-03-27 & 08:16:28.01 & VLT & SPHERE/ZIMPOL & R & -410.5 & 2.8 & 0.9 & 1.3 & 3.6 & 9.0 & 0.3  \\ 
2019-04-10 & 07:20:44.49 & VLT & SPHERE/ZIMPOL & R & 383.1 & -20.2 & -0.6 & 0.8 & 3.6 & 8.8 & 0.2  \\ 
  \end{longtable}
\end{center}
\twocolumn

\end{appendix}

\end{document}

%% file: tab_ao.tex
\begin{table*}
\caption{\label{tab:ao}List of VLT/SPHERE disk-resolved images obtained in the I filter by the ZIMPOL camera. For each observation, the table gives the epoch, the exposure time, the airmass, the distance to the Earth $\Delta$ and the Sun $r$, the phase angle $\alpha$, and the angular diameter $D_\mathrm{a}$.}
\centering
\begin{tabular}{rr rr rrr r}
\hline 
\multicolumn{1}{c} {Date} & \multicolumn{1}{c} {UT} & \multicolumn{1}{c} {Exp} & \multicolumn{1}{c} {Airmass} & \multicolumn{1}{c} {$\Delta$} & \multicolumn{1}{c} {$r$} & \multicolumn{1}{c} {$\alpha$} & \multicolumn{1}{c} {$D_\mathrm{a}$} \\
\multicolumn{1}{c} {} & \multicolumn{1}{c} {} & \multicolumn{1}{c} {(s)} & \multicolumn{1}{c} {} & \multicolumn{1}{c} {(AU)} & \multicolumn{1}{c} {(AU)} & \multicolumn{1}{c} {(\degr)} & \multicolumn{1}{c} {(\arcsec)} \\
\hline\hline
  2019-03-15 &      7:07:59 & 183 & 1.07 & 2.33 & 3.18 & 10.6 & 0.159  \\
  2019-03-15 &      7:11:12 & 183 & 1.07 & 2.33 & 3.18 & 10.6 & 0.159  \\
  2019-03-15 &      7:14:25 & 183 & 1.07 & 2.33 & 3.18 & 10.6 & 0.159  \\
  2019-03-15 &      7:17:37 & 183 & 1.07 & 2.33 & 3.18 & 10.6 & 0.159  \\
  2019-03-15 &      7:20:49 & 183 & 1.07 & 2.33 & 3.18 & 10.6 & 0.159  \\
  2019-03-20 &      4:16:29 & 183 & 1.35 & 2.30 & 3.19 & 9.1 & 0.161  \\
  2019-03-20 &      4:19:43 & 183 & 1.33 & 2.30 & 3.19 & 9.1 & 0.161  \\
  2019-03-20 &      4:22:56 & 183 & 1.32 & 2.30 & 3.19 & 9.1 & 0.161  \\
  2019-03-20 &      4:26:07 & 183 & 1.31 & 2.30 & 3.19 & 9.1 & 0.161  \\
  2019-03-20 &      4:29:20 & 183 & 1.29 & 2.30 & 3.19 & 9.1 & 0.161  \\
  2019-03-25 &      2:53:48 & 183 & 1.73 & 2.27 & 3.20 & 7.5 & 0.163  \\
  2019-03-25 &      2:57:01 & 183 & 1.71 & 2.27 & 3.20 & 7.5 & 0.163  \\
  2019-03-25 &      3:00:14 & 183 & 1.68 & 2.27 & 3.20 & 7.5 & 0.163  \\
  2019-03-25 &      3:03:26 & 183 & 1.65 & 2.27 & 3.20 & 7.5 & 0.163  \\
  2019-03-25 &      3:06:38 & 183 & 1.62 & 2.27 & 3.20 & 7.5 & 0.163  \\
  2019-03-27 &      8:16:28 & 183 & 1.23 & 2.26 & 3.21 & 6.8 & 0.164  \\
  2019-03-27 &      8:19:41 & 183 & 1.24 & 2.26 & 3.21 & 6.8 & 0.164  \\
  2019-03-27 &      8:22:55 & 183 & 1.25 & 2.26 & 3.21 & 6.8 & 0.164  \\
  2019-03-27 &      8:26:06 & 183 & 1.27 & 2.26 & 3.21 & 6.8 & 0.164  \\
  2019-03-27 &      8:29:18 & 183 & 1.28 & 2.26 & 3.21 & 6.8 & 0.164  \\
  2019-04-08 &      3:58:12 & 184 & 1.13 & 2.24 & 3.23 & 2.8 & 0.165  \\
  2019-04-08 &      4:01:27 & 184 & 1.13 & 2.24 & 3.23 & 2.8 & 0.165  \\
  2019-04-08 &      4:04:40 & 184 & 1.12 & 2.24 & 3.23 & 2.8 & 0.165  \\
  2019-04-08 &      4:07:52 & 184 & 1.12 & 2.24 & 3.23 & 2.8 & 0.165  \\
  2019-04-08 &      4:11:06 & 184 & 1.11 & 2.24 & 3.23 & 2.8 & 0.165  \\
  2019-04-10 &      2:30:41 & 184 & 1.36 & 2.24 & 3.24 & 2.3 & 0.165  \\
  2019-04-10 &      2:33:56 & 184 & 1.35 & 2.24 & 3.24 & 2.3 & 0.165  \\
  2019-04-10 &      2:37:09 & 184 & 1.34 & 2.24 & 3.24 & 2.3 & 0.165  \\
  2019-04-10 &      2:40:23 & 184 & 1.32 & 2.24 & 3.24 & 2.3 & 0.165  \\
  2019-04-10 &      2:43:36 & 184 & 1.31 & 2.24 & 3.24 & 2.3 & 0.165  \\
  2019-04-10 &      7:20:44 & 184 & 1.27 & 2.24 & 3.24 & 2.2 & 0.165  \\
  2019-04-10 &      7:23:57 & 184 & 1.29 & 2.24 & 3.24 & 2.2 & 0.165  \\
  2019-04-10 &      7:27:12 & 184 & 1.30 & 2.24 & 3.24 & 2.2 & 0.165  \\
  2019-04-10 &      7:30:24 & 184 & 1.31 & 2.24 & 3.24 & 2.2 & 0.165  \\
  2019-04-10 &      7:33:38 & 184 & 1.32 & 2.24 & 3.24 & 2.2 & 0.165  \\
\hline
\end{tabular}
\end{table*}